# Multifunctional Control of On-chip Generated Photons by a Single Collective Mode in Monolithically Integrated All-Dielectric Scalable Optical Circuits

SWARNABHA CHATTARAJ,[1] JIEFEI ZHANG,[2] SIYUAN LU,[3] AND ANUPAM MADHUKAR[*,1,2,4]

[1]*Ming Hsieh Department of Electrical Engineering, University of Southern California, Los Angeles, California 90089, USA*
[2]*Department of Physics and Astronomy, University of Southern California, Los Angeles, California 90089, USA*
[3]*IBM Thomas J. Watson Research Center, Yorktown Heights, New York, 10598, USA*
[4]*Mork Family Department of Chemical Engineering and Materials Science, University of Southern California, Los Angeles, California 90089, USA*
**\*** *madhukar@usc.edu*

**Abstract:** Recently we proposed the use of a single collective Mie mode of coupled dielectric resonators arranged in metastructures designed to simultaneously enhance the emission rate, control the direction, and propagation of emitted photons from an embedded single photon source (SPS) [J. Opt. Soc. Am. B 33(12), 2414(2016); J. Appl. Phys.120, 243103(2016)]. Here we extend the analysis to include the necessary additional two functions of beam splitting and controlled combing of photons from two distinct on-chip sources for interference. Thus, the same single Mie mode of the designed system provides all the required five functions in different spatial regions of an on-chip integrated optical circuit, enabling full control over photon interference for classical and quantum optical information processing. Single photon source architectures suited for such on-chip scalable integration are discussed.

## 1. Introduction

Realization of on-chip optical circuits that allow controlled interference between photons is a critical milestone to be reached towards optical information processing in both classical and quantum regimes. Photons have advantages as information carrier as they enable long distance and near-lossless fast communication. Entanglement between a matter-based qubit (e.g. the electron spin in a quantum dot) and propagating photonic qubit has recently been achieved [1] towards more recent demonstration of photon-mediated entanglement between spin qubits [2]. Moreover, photons also allow realization of quantum information processing by utilizing linear optical circuits that only comprise beam-splitters and phase shifters, as first proposed in ref [3].

Since reference 3, the potential of single photon based optical circuits for quantum information processing has been investigated in the platform of silicon photonics where silicon waveguide structures seamlessly changing into beam-splitters and subsequently beam-combiners resulting in potentially very high-fidelity quantum logic gate implementation [4-7]. However, not only do optical circuits based on such conventional waveguide technology demand large on-chip footprint of ~10um-100 um scale, but they are also severely limited by the unavailability of on-chip integrable single photon sources (SPSs). So far, the SPSs in these implementations are mostly realized by (1) exploiting highly inefficient spontaneous-four-wave-mixing in Si with the help of an external pumping laser [4-7], or (2) in some cases, hybrid integration with III-V based island quantum dots [8, 9]. Neither of these implementations allow scalability. Monolithic integration of on-chip SPSs with suitable light manipulating optical circuits is thus of paramount importance towards realization of fully self-sufficient and scalable nanophotonic systems.

Considerable challenges, however, are faced in realizing monolithically integrated systems as they require that the on-chip SPSs be in spatially regular array with precise control on spectral and spatial matching between the emitted photon and the modes of the constituents (cavity, waveguide, etc.) of the on-chip optical circuit. The required five functions constitute (1) deterministic efficient coupling of the emitter with a resonant cavity to enhance emission rate, (2) efficient directed escape into a waveguide, (3) state-preserving propagation, (4) splitting, and (5) combining photons emitted from different sources to achieve controlled on-chip photon interference. So far, the dominant approach towards this long-sought goal has been coupling of compound semiconductor (typically the strained InGaAs-GaAs system) self-assembled island quantum dot [10, 11] based SPSs with a cavity and / or waveguide, hereafter referred to as light manipulating elements (LMEs), implemented predominantly using 2D photonic crystal membrane structures [12-14]. In this approach, point or line defects in an otherwise appropriate designed array of holes in defining the photonic 2D crystal in a semiconductor membrane architecture are utilized to create localized photon modes based on departures from Bragg scattering to provide, so far typically, resonant cavity function [15-17] or/and wave-guiding function (typically each designed independently) [18]. The random spatial distribution and inherent wide spectral emission characteristics of the SAQD SPSs however have prevented examination of optical circuits in such systems. The studies of site-selected single quantum dot (SQD) SPS synthesized in arrays of pre-patterned holes in the substrate pioneered by the Kapon group [19, 20] are an exception and naturally suited for on-chip integration with optical circuits fabricated in the surrounding dielectric medium (here GaAs). Realization of a fully functional complete optical circuit providing the above noted five required functions in the current photonic 2D crystal approach has not been reported. Indeed, implementations of individual functions of a cavity and / or waveguide around a SQD source are in considerable need of improved spectral and spatial overlap matching of the emitter wavelength and the modes of the cavity and the waveguide.

To address these challenges, recently we introduced a new approach [21, 22] to realizing on-chip classical and quantum optical networks for manipulating photons emitted, on-demand, from spatially regular arrays of spectrally uniform SPSs [22-25]. In this approach the light manipulating functions are implemented through a single collective Mie resonance of co-designed planar metastructures made of sub-wavelength sized dielectric building blocks (DBBs). The metastructures, hereafter called light manipulating units (LMUs), provide, *simultaneously,* the needed five light manipulating functions. Figure 1(a) schematically depicts such an envisioned on-chip integrated optical network in which each single photon source (depicted as point-like sources shown in red buried in appropriate media) in an array is coupled to the DBB metastructure (blue blocks) that extracts the emitted photons in a horizontal direction and guides, propagates, and interferes photons from distinct SPSs on-chip. In this paper we present the results of simulations of a SPS-DBB integrated complete optical circuit (Fig. 1(b)) that provides all these required light manipulating functions.

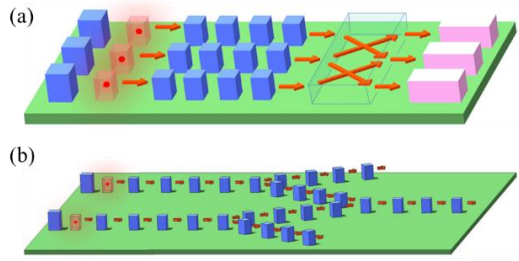

Fig. 1. (a) The envisioned new approach [18] to quantum optical circuits comprising array of single photon sources (represented by the red dots) integrated with co-designed multifunctional light manipulating structures based on sub-wavelength sized dielectric building blocks (DBBs, blue blocks) and single photon detectors (purple blocks). (b) Schematic representation of a DBB array based optical network that exploits a single collective Mie resonance of the interacting DBBs to achieve the required five functions: Purcell enhancement and directed-emission (nanoantenna effect), wave-guiding, beam-splitting, and beam-combining, thus enabling interference between single photons from distinct SPSs in an optical circuit.

The DBB metastructure paradigm exploits the existence of strong magnetic resonances in the optical wavelength regime in sub-wavelength sized *high refractive index* dielectric building blocks arranged in engineered architectures, such as that of Fig. 1(b), to tailor their collective Mie resonance in both its spatial function and spectral nature by adjusting the size, shape, and configuration of the building blocks. We note that, exploiting Mie resonance in subwavelength size dielectric building block linear arrays (as contrasted with the conventional continuous medium slab waveguide) for wave guiding light in the sub-diffraction length scale has been under examination exploiting suitable low-order dipole and multipole modes with strong magnetic response [26-31]. The simulations of implementation of *all five functions* needed for manipulating on-chip generated photons to achieve functional optical circuits reported here are, however, a first. As will be seen below, this enables reducing footprint of the on-chip optical circuits [27] for implementation of the waveguiding and beamsplitting structures and also helps integration with photon sources utilizing nanoantenna structures [28]. Recently the use of Mie resonance has been explored experimentally for implementation of low loss waveguides [31] and enhancement of photoluminescence from embedded island QDs [32]. However, at this early stage these theoretical and experimental explorations have been limited to individual functions. Needed are investigations of fully functional quantum optical circuit and in the following we report the simulated behavior of an optical circuit.

The novelty of the architectures that we report on here lies in the fact that the same single collective Mie resonance of the whole designed structure (dubbed light manipulating unit or LMU) provides *all* the required five functions in a quantum optical circuit: (1) enhancement of the radiative decay rate of the SPS through a local enhancement of the electric field (the Purcell effect), (2) imposing directionality on photon emission (nanoantenna effect), (3) state-preserving propagation with controlled phase shift of the single photons (waveguiding), (4) splitting the single photon on-chip (beam splitting), and (5) merging two distinct photons onto the same branch (beam-combining) to create interference. Photons in such an optical circuit seamlessly propagate in the same collective optical mode that in different spatial regions of the circuit performs different functions. Unlike conventional architectures that have so far implemented integration of different functions via individual components (cavity, waveguide) each with its own characteristic mode which brings into play considerations of mode-matching, the Mie resonance approach does not create such concern—it reduces the design challenge to only one matching problem, i.e., spectral and spatial mode matching between the electronic state of the SPS and the single collective Mie resonance of the whole optical circuit.

Deterministic spectral and spatial matching between the electronic state of the SPS and the light manipulating circuits, however, has been proven to be a challenging task. Traditionally the self-assembled island quantum dots have dominated as single photon source [10-14] but their inherently random spatial location and spectral emission prevent fabrication of optical circuits. In recent years deep level defect centers (e.g. NV and SiV center in diamond) have been established SPSs with very good spectral uniformity and, via focused ion beam implantation, embedded with ~50nm precision [33-35]. However, the low (~10%) yield of the ion implantation process makes controlling the number of SPS per site very challenging. To realize spatially regular array of SPSs, single quantum dots have been grown in spatially patterned arrays of recesses [19, 20] and, recently on the top of engineered mesa arrays [22-24]. The latter, new class of SQDs dubbed mesa-top SQDs (MTSQDs), are formed on the apex of lithographically defined mesas and have been demonstrated to act as SPSs with more than 90% purity. Furthermore, the MTSQDs can be readily overgrown and planarized [22] thereby facilitating lithographic integration with the DBB array based optical circuit as envisioned in Figure 1(a). With such spectrally tuned and spatially regular MTSQD SPS array readily imbedded in DBB based LMUs, the MTSQD-LMU interconnected structures can be realized to create optical circuits that enable on-chip interference, enabling a first step towards quantum information processing.

## 2. The DBB based Nanoantenna-Waveguide Unit with Mie Theory

To gain physical insight into the collective multipole Mie modes through calculations based on analytical expressions of the Mie theory, we model the DBBs as spheres (Fig.2a). Shapes that lend themselves to lithographic fabrication can be readily simulated using numerical analyses such as FEM and FDTD. The spherical shape allows the expansion of the fields in the basis of multipole modes described as spherical vector harmonics – also known as the TE (transverse electric) and TM (transverse magnetic) modes.

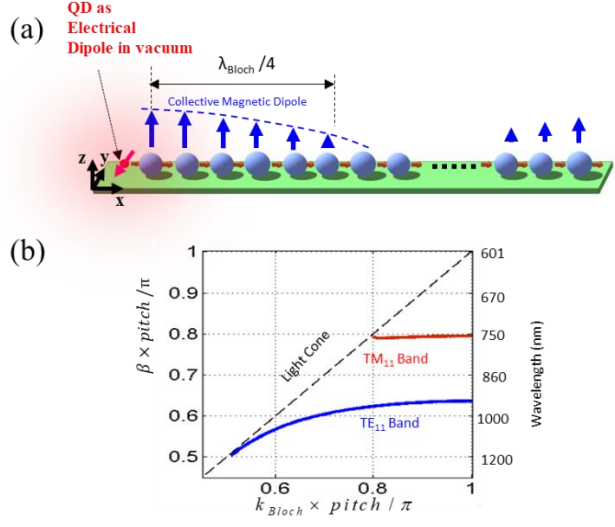

Figure 2. (a) schematic showing an array of interacting spherical DBBs excited by a point electrical dipole representing an on-chip photon emitter. The collective magnetic mode is schematically represented by the blue arrows. (b) The calculated dispersion characteristics of the collective Mie mode of an array of DBBs of radius 129nm, refractive index 3.5 and pitch 300.6nm.

Following this approach, the electromagnetic fields of the DBB array can be expressed as a superposition of different TM and TE modes as the following [21, 22, 36]:

$$\bar{E}(\bar{r}) = \sum_{i=1}^{N} \sum_{n=1}^{n_{max}} \sum_{m=-n}^{n} \left( a_{n,m}^{(i)} \bar{E}_{TEn,m}(\bar{r} - \bar{r}_i) + b_{n,m}^{(i)} \bar{E}_{TMn,m}(\bar{r} - \bar{r}_i) \right) \ldots \ldots (1)$$

Here $\{a_{n,m}^{(i)}, b_{n,m}^{(i)} \mid i = 1:N; n = 1:n_{max}; m = -n:n\}$ denote the coefficients corresponding to the different multipoles of the constituent DBBs of the optical network comprising $N$ DBBs. In an interacting array of DBBs, the modes of the individual DBBs can mix, resulting in a collective mode that is spatially delocalized over the whole array. This delocalized collective Mie mode can, indeed, be coupled to an on-chip photon generator. This is schematically illustrated in Fig. 2(a), where a point electric dipole (representing an on-chip localized photon source) is coupled to a linear-chain of DBBs, and therefore excites the collective magnetic dipole mode of the array, depicted using the blue arrows. To identify the nature of this coupled mode, we performed analysis of the dispersion characteristics for a specific case of GaAs DBBs of radii 129nm— targeted for the collective magnetic dipole resonance at the desired operating wavelength of 980nm (see Appendix 1). The results, as shown in Fig. 2(b) reveal the existence of bands with real wave-vector $k_{Bloch}$ that propagates the coupled light losslessly along the array. Also, the Bloch nature of the collective mode in the linear array allows to impart phase shift on the propagated light proportional to the propagation length. The phase shift as a function of the propagation length L is thus simply given by $k_{Bloch} \times L$. Furthermore, since the bands do not overlap, the collective magnetic dipole mode at 980nm does not have any degeneracy and essentially acts as a single mode — a fact that is well exploited in the design different components of the optical circuits presented next.

**Nanoantenna-waveguide: Physics of the nanoantenna**

As shown in Fig. 3(a), the nanoantenna unit [22] is realized by placing the reflector DBB on the other side of the array (shown in purple). Similar to the well-known Yagi-Uda architecture, the presence of the reflector along with the waveguiding DBBs enhance the directionality of the emitted light from the source dipole. This is calculated by computing the angular E-field distribution on a spherical surface centered on the source dipole. The resultant angular E-field pattern is shown in the inset of Fig.3(a). Integrating the angular fields over the cross section of the waveguide section indicate a coupling efficiency of ~60% of the emitted light from the source dipole to the collective magnetic dipole mode of the array. The coupled photons are propagated without loss, as indicated by the behavior of the $a_{1,1}$ coefficients (representing the magnetic dipole mode) shown along the DBB array in Fig. 3(c). The same collective mode also enhances the E-field at the location of the emitter, resulting in enhancement of the photon local density of states at the location of the source dipole. This is estimated using the Green function based approach [21], revealing an estimated a Purcell enhancement of ~7 (Fig. 3(d)) at the targeted operating wavelength of 980nm that will shorten the radiative lifetime of the SPSs and therefore improve the photon indistinguishability enabling photon interference [37, 38]. The Purcell enhancement is facilitated by the Fabry - Perot interference effect of the finite segment of the nanoantenna-waveguide section, as pointed out in ref.[22]. In fact, near the band edge, the highly dispersive nature of the propagating wave allows to obtain a Purcell enhancement even higher [22]. Our design, however, is chosen to allow a ~4nm width to the peak at 980nm shown in Fig.3(d). This is motivated to account for the current spectral uniformity of the MTSQD based optical circuits.

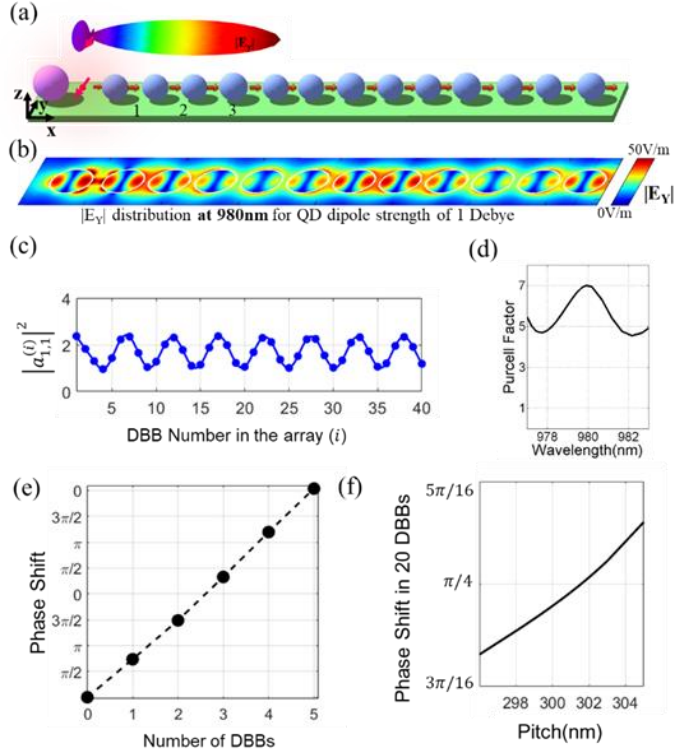

Fig. 3. (a) A schematic showing the nanoantenna-waveguide structure comprising GaAs DBBs (blue spheres) embedded with a SPS (yellow pyramid) modeled as a dipole emitter. The angular distribution of the E-field is also shown, indicating the enhancement of the directionality of the SPS in the presence of the DBBs. The DBBs in the array are numbered 1 to 40 for quick reference. (b) E-field distribution of the collective magnetic dipole mode of the array. (c) The expansion coefficient of the magnetic dipole along the array of 40 DBBs, indicating lossless propagation over the whole distance. The collective magnetic dipole mode enhances the E-field at the position of the SPS, causing enhancement of the emission rate (Purcell factor) as shown in Panel (d) (From Ref. [22]). (e) Shift in the phase of the magnetic dipole mode ($\angle a_{1,1}^{(i)}$), representing the phase shift of the propagating photon as a function of the number of DBBs for an array of pitch 300.6nm—showing linear dependence of the phase shift with propagation distance. (f) The phase shift at 980nm by a segment of 20 DBBs as a function of the pitch of the array.

The phase-shift in the nanoantenna-waveguide system is illustrated in Fig. 3(e)—showing linear dependence of the phase-shift with propagation distance. This is caused by the Bloch nature of the propagating field distribution of the array of the DBBs that locally act like linear chain of coupled resonators [39]. Additionally, such phase shift can be finely adjusted by changing the pitch of the DBB array. For example, in Figure 3(f), we show that the phase shift by a segment of 20 DBBs can be tuned to a value of $\frac{\pi}{4}$, that has great significance in proposed implementations of quantum information processing [3]. This also opens the possibility of dynamically changing the phase of the photons by methods such as changing the free carrier density by electrical injection or photoexcitation [40-42].

In addition to the enhanced and directed emission and guiding and phase-shifting of the photons, a complete optical circuit demands control on bifurcating the photon propagation pathway (i.e. beam-splitting) and combining photons from two different propagating pathways. To this end, we present next the new results of our investigations of the complete nanoantenna-waveguide-beam splitting and beam-combining LMUs. We emphasize that all functions are provided by the same single collective Mie resonance of the designed structure to eliminate mode-mismatch issues amongst different parts of the optical circuit and enable control on photon interference in an on-chip platform.

### 3.  Nanoantenna - Waveguide - Splitter - Combiner for photon interference

In Figure 4 we show schematic of a complete optical circuit that provides the beam-splitting and beam-combining functionalities along with the enhancement of the emission rate of the photon source enhancement in directionality, and lossless propagation of the emitted photons with controlled phase shift. Such an optical circuit is both necessary and sufficient to understand and assess the functioning of a SPS-LMU on-chip integrated information processing system. To facilitate analysis of such a multifunctional system based on Mie theory, we model the DBBs as spheres (Fig. 4) which, as noted previously, allows the expansion of the fields in the basis of multipole modes described as spherical vector harmonics as noted above. A shape more consistent with lithographic fabrication, such as rectangular as shown in Fig. 1(b), can be simulated using more computationally demanding numerical tools such as the finite element method or the finite-difference time-domain method. The general approach of exploiting a single collective mode of a DBB array to implement multiple functions, however, remains independent of the shape of the resonator.

The results presented below are for a LMU structure (Fig. 4) that is designed as a natural extension of the nanoantenna-waveguide unit [22] to include beam-splitting and combining. The simulated circuit of Fig. 4 comprises spherical GaAs DBBs of radius 129nm arranged in an array with center-to-center spacing of 300.6nm. This results in the collective magnetic dipole resonance of the array to be centered around the targeted emission wavelength of 980nm of the photon source. The photon source is modelled as a point electrical dipole with a polarization in-plane, along the y direction. The purple DBB of radius 131.7nm is placed opposite to the array to act as the reflector to provide the nanoantenna function, as discussed in Ref [22]. We find that in the simulated results of the complete circuit the addition of the beam-splitting and beam-combining functions does not affect the functions of nanoantenna-waveguide piece. The Purcell enhancement, enhancement of directionality of photon emission, and the lossless propagation and phase shift in the complete nanoantenna-waveguide-beam splitter - beam combiner circuit (Fig.4) remain essentially same as the results reported in [22] and recalled here in Fig.3 (b)-(d). Thus, in the discussion below we focus only on the findings for the beam-splitting and the beam-combining operations that lead to photon interference.

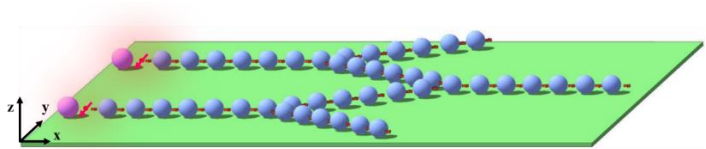

Fig. 4. The complete structure of the nanoantenna-waveguide-beam splitter-beam combiner unit whose single mode - the collective magnetic dipole Mie resonance - provides all the five needed on-chip light manipulating functions: enhancement of the emission rate of the SPSs (shown as red arrows representing the source dipole), imposing directionality on emission, state-preserving wave-guiding, photon path splitting, and photon combining.

*Beam Splitting*

The splitting of the photons is achieved using the same collective Mie mode as the propagation, i.e. the collective magnetic dipole mode of the DBB array. The beam-splitting structure is implemented by creating a fork comprising two nearest neighbors of the DBB waveguide segment as depicted in Fig. 5(a) with a center-to-center separation of ~300.6nm, same as the waveguide structure. This results in the magnetic dipole mode of the DBB segment at the junction coupling to the magnetic dipole modes of the two starting DBBs of the two branches at the fork, creating a collective Mie resonance mode that splits the energy between the two branches equally. The resulting collective magnetic dipole mode is shown via its E-field distribution in Fig. 5(b). The main nanoantenna-waveguide structure consists of 20 DBBs, whereas each individual branch of the beam-splitter consists of 20 DBBs each. An angle of $60^0$ is taken between the two branches in the x-y plane. Fig. 5(c) shows the magnetic dipole mode coefficient as a function of the DBB number in the wave-guiding segment. An oscillatory, but non-diminishing, behavior indicating lossless propagation is revealed. By design, the splitting between the two branches is symmetric. It thus suffices to show in Fig. 5(d) the magnetic dipole mode coefficient for only one of the individual branches after the splitting.

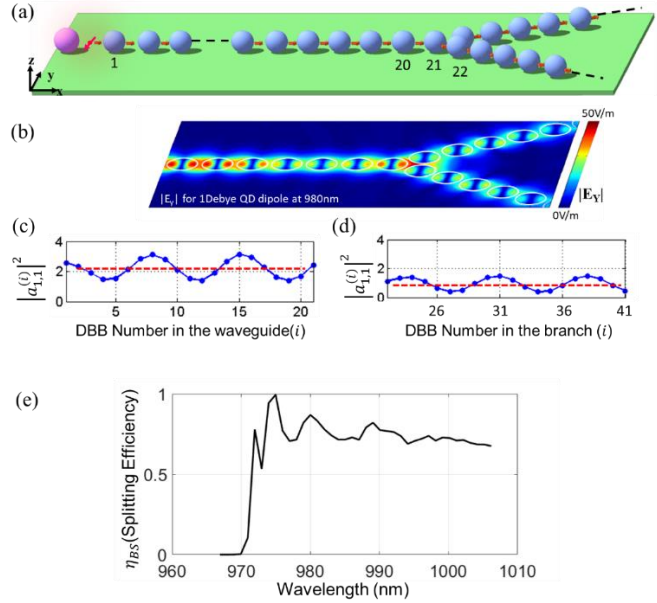

Fig. 5. (a) Schematic showing up to the beam-splitting component of the optical circuit of Fig.4. For ready reference in the subsequent plots, the DBBs are numbered as shown. (b) The E-field distribution of the collective magnetic dipole resonance of the complete structure. (c) and (d) show the magnitude of the magnetic dipole mode coefficient along the main waveguide before splitting, and along one of the branches after splitting, respectively. The red dashed line represents the average value of $\left|a_{1,1}^{(i)}\right|^2$ along the array. (e) The spectrum of the beam-splitting efficiency ($\eta_{BS}$, see text).

From Fig. 5(c) and Fig. 5(d), we also observe that the magnetic dipole mode amplitude does not remain constant over the array and shows oscillatory behavior. This oscillation originates from Fabry-Perot resonance and the oscillation period signifies the Bloch wave-vector of the collective magnetic dipole mode of the array. To estimate the average energy of the collective mode in each branch, we thus compute the average value of the amplitude square of the magnetic dipole coefficient, denoted by the red dashed lines in Fig. 5(c) and Fig. 5(d). The splitting is achieved by the resonant coupling of the magnetic dipole mode of the DBB of the main waveguide section with the magnetic dipole modes of the DBBs of the two branches. We define the efficiency of such a coupling process resulting in splitting of the propagating photon as the ratio of the average value of magnetic dipole mode intensity of the two branches divided by the average magnetic dipole mode intensity in the main waveguide before the splitting, i.e.,

$$\eta_{BS} = 2 \times \frac{avg|a_{1,1}(Branch)|^2}{avg|a_{1,1}(Main)|^2} \ldots\ldots\ldots\ldots\ldots\ldots\ldots\ldots\ldots\ldots\ldots\ldots. (2)$$

The splitting efficiency is plotted in Fig. 5(e) as a function of wavelength. We observe that a high beam-splitting efficiency of near unity is readily achievable. The slight deviation from unity efficiency may be attributed to scattering loss in the junction owing to the broken translation symmetry of the DBB array. Moreover, the splitting efficiency remains fairly large (>0.75) over a large wavelength range of ~970nm -1000nm, i.e. the whole band of the collective magnetic dipole mode [21, 22]. We note that such robust behavior is contributed by the fact that the same collective magnetic dipole mode is exploited for the splitting which ensures spatial mode-matching over the whole wavelength range. This will also prove to be an advantage in designing hierarchically more complex optical network using this approach.

*Beam Combining and Photon Interference:*

Following splitting of the beams originating from two photon sources (SPS1 and SPS2) as depicted in Fig. 6(a), combining the photons (red and magenta) from the two sources with controlled phase separation is the key operation in the optical circuit. Thus, we next present the results of the simulations of beam-combining function achieved by coupling the two branches onto a single branch exploiting the coupling between the magnetic dipole modes of the constituent DBBs.

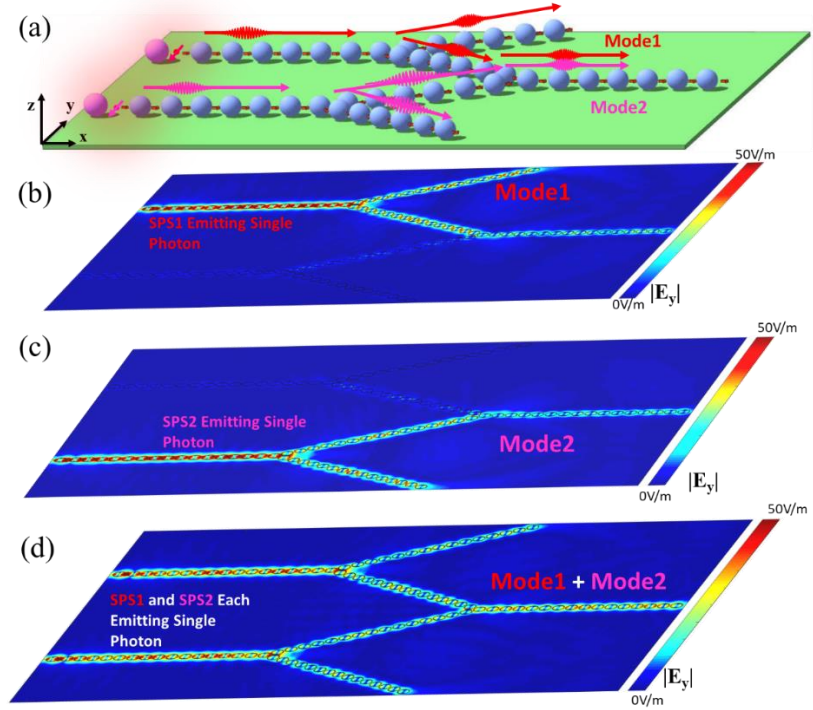

Fig. 6. (a) The complete nanoantenna-waveguide-beam splitting- beam combining unit. The emission from the two SPSs are coupled to the two collective magnetic dipole mode represented by the red (Mode1) and magenta (Mode2) arrows. Panels (a) and (b) show, respectively, the electric field distribution for the red and magenta beam splitting pathways. Panel (d) shows the electric field distribution of the two-photon state when the two SPSs emit simultaneously.

Note that fundamentally there are two single collective magnetic dipole modes of the entire unit that play a role in beam-recombining and photon interference. These two collective modes, depicted as red and magenta arrows respectively in Fig. 6(a), coupled to the two distinct SPSs, provide *simultaneously* the Purcell enhancement, nanoantenna, waveguide and splitting functions as discussed before. The two modes, however, finally overlap in the central branch, allowing the two photons to interfere. As an illustration, Fig. 6(b) shows the electric field distribution when only SPS1 has emitted a photon into Mode 1 (collective magnetic dipole mode in the top branch) of the unit. Equivalently, when SPS2 only has emitted a photon in Mode 2 (collective magnetic dipole mode in the bottom branch) of the unit, the electric field distribution is shown in Fig. 6(c). The interference takes place when both SPS1 and SPS2 emit simultaneously. The electric field distribution for such a two-photon state is found by superposition of the electric fields produced by the individual SPSs. Such combined electric field distribution corresponding to the in-phase interference of the two photons is shown in Fig. 6(d), illustrating the interference of the two photons in the common branch of the beam-combining unit. Other states with a non-zero phase difference between the two interfering photons can be readily achieved by independently adjusting the phase shifts in the two nanoantenna-waveguide branches, as discussed in detail in Section 2. Dynamical control over the phase shift of the photons may also be achieved by tuning the Mie resonance frequencies of the DBBs via dynamic control on the refractive index with carrier injection techniques [40-42].

The interference of two distinct spectrally matched single photons from the two sources onto the single collective mode of the combined branch will also enable entanglement between the two sources. Such entanglement is created by the act of detecting a single photon in the collective mode of the combined branch—as the information of the origin of the photon is lost. Thus, a single photon measurement event on the combined branch will project the two SPSs onto a path-entangled state. Interestingly, such entangled state shows super-radiance— an identity that can be used to experimentally verify entanglement in such systems [35]. Therefore, the realization of SPS-LMU can enable the first step towards realizing quantum optical circuits. Further experimental work on fabricating such structures is underway.

## 4. Conclusion

The work reported here contributes to realizing the goal of on-chip integrated nanophotonic systems for classical and quantum information processing, shown schematically in Fig. 1(a). Our approach comprises spatially regular array of single photon sources (SPSs) embedded in *co-designed* light manipulating linear optical networks that control emission rate, direct, guide the emitted single photons with controlled phase shift, and enable interference of single photons from distinct SPSs, thus satisfying the needed functions towards on-chip optical quantum information processing.

Towards this goal, recent advance in the realization of single photon sources based on a novel class highly spectrally uniform semiconductor single quantum dots dubbed mesa-top single quantum dot (MTSQD) grown in *spatially regular arrays* essential for enabling optical networks [22] provide renewed incentive for their integration with light manipulating units. For the LMUs we proposed [21, 22] an alternative to the traditional practice of using the 2D photonic crystal platform for implementation of emitted photon manipulating elements (resonant cavity, waveguide, etc.) that potentially provides comparable functions but at a much reduced on-chip footprint, thus also enabling higher scalability. The new approach to implementing the requisite multiple functions for realizing an optical network is fundamentally different from the departure from Bragg diffraction mechanism underlying the conventional photonic crystal approach. It exploits a single collective Mie resonance of an array of subwavelength sized interacting high refractive index dielectric building blocks arranged around each SPS in metastructures (blue blocks, Fig.1) co-designed to provide *all* the needed multiple functions: enhancement of the emission rate of the SPS, enhancement of the directionality of emission, on-chip state-preserving propagation with controlled phase shift, beam-splitting and beam-combining functions. Apart from the promising approach of MTSQD SPSs, the proposed approach of optical circuits based on DBB array exploiting a single collective Mie mode is equally compatible with other regular arrays of SPSs such as defect centers, that satisfy the basic requirement of spectral and spatial matching between the electronic state and the photonic mode of the single collective Mie resonance.

In this paper we have reported simulations of the behavior of photons generated by a point dipole source embedded in a DBB based multifunctional light manipulating unit whose collective dipole mode, primarily magnetic, at the designed SQD emission wavelength of 980nm, provides the above noted five functions in different spatial regions of the unit acting as nanoantenna-waveguide-beam splitter-beam combiner to build optical circuit as shown in Figure 4. A Purcell enhancement of ~7 at the targeted emission wavelength of 980nm is obtained, adequate to enable demonstration of indistinguishable single photon emission from the MTSQDs and observation of photon interference from distinct on-chip sources in a scalable architecture. State preserving lossless propagation with a $\pi/4$ phase shift on the photons is demonstrated, satisfying a critical requirement for photon-based quantum information processing [3]. Furthermore, beam splitting with near unity efficiency and combining emitted photons from two distinct SPSs to the same collective Mie resonance -- resulting in interference - - is presented.

To gain physical insight, we have employed analytical expressions of the Mie theory for spherical shape dielectric building blocks to arrive at the presented results. However, the physical properties of the collective Mie resonance of such multifunctional optical circuits are equally relevant to DBBs of non-spherical shape. As depicted in the envisioned optical circuit in Figure 1 (a) and 1(b), compatibility with on-chip SPSs in terms of monolithic integration requires lithographic fabrication of the DBBs and thus rectangular or cylindrical shaped DBBs. We have recently undertaken numerical simulation based on COMSOL, Finite element method to design and optimize such optical circuits comprising rectangular DBBs to guide the current efforts to fabrication of such structures [25]. However, irrespective of the actual shape of the DBBs, the approach of exploiting a single Mie resonance of the whole structure eliminates the need for impedance matching between different functional parts of the optical circuit as all the needed light manipulating functions are delivered by the same collective mode. Such an approach allows implementation of more complex optical networks built from the same fundamental functional elements as demonstrated in this paper. Such SPS-DBB integrated networks, we believe, are promising candidates deserving full exploration towards optical quantum information processing systems.

**Funding**

This work is supported by Army Research Office (Program Manager Dr. John Prater), Grant Number: W911NF-15-1-0298.

**Appendix 1: Mathematical Methods -- Mie Theory**

Here we elaborate on the Mie theory-based analysis that is used to calculate the response of the array of DBBs of spherical shape—under the excitation of an arbitrary point or distributed classical source. The basic conceptual picture is shown in Figure A1. For elaboration, an arbitrary ensemble of spherical shaped N number of scatterers are shown, with their center positions being $\{\bar{r}_1, \bar{r}_2, \bar{r}_3, ... \bar{r}_N\}$, and radii being $\{R_1, R_2, .... R_N\}$. The assembly of scatterers is illuminated by an arbitrary source as depicted by the red wave front. Under weak-coupling of the scattered wave with the source, one can express the total electric field at any arbitrary point P as

$$\bar{E}_{Total}(\bar{r}) = \bar{E}_{Source}(\bar{r}) + \bar{E}_{Scattered}(\bar{r}). \ldots\ldots\ldots\ldots\ldots\ldots\ldots\ldots\ldots\ldots\ldots(A1)$$

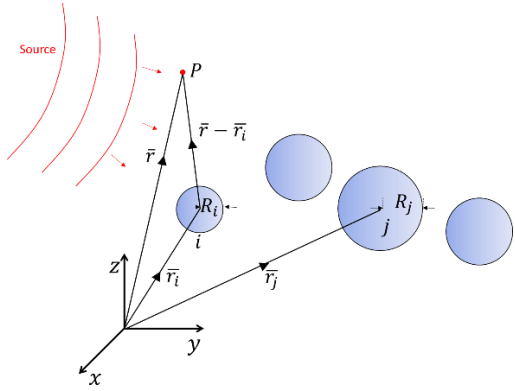

Figure A1. Schematic showing an array of non-overlapping spherical DBBs excited by an arbitrary source (red wavefront).

The same superposition relation holds for any electromagnetic field quantity such as the magnetic field and the electric and magnetic vector potential. Now, the scattered wave can be thought of as originating from the center of each spherical scatterers and therefore can be expressed in the basis of the spherical vector harmonics around each point. Thus, we have

$$\bar{E}_{Scattered}(\bar{r}) = \sum_{i=1}^{N} \sum_{n=1}^{n_{max}} \sum_{m=-n}^{n} \left( a_{n,m}^{(i)} \bar{E}_{TEn,m}(\bar{r} - \bar{r}_i) + b_{n,m}^{(i)} \bar{E}_{TMn,m}(\bar{r} - \bar{r}_i) \right) \ldots \ldots (A2)$$

The spherical vector harmonics, expressed in the form of TE (radially transverse electric field mode) and TM (radially transverse magnetic field mode) can be analytically expressed as

$$\bar{E}_{TEnm}(\bar{r}) = \frac{1}{\sqrt{n(n+1)}} \left( in z_n(\beta r) \frac{Y_{n,m}(\theta,\phi)}{\sin(\theta)} \hat{\theta} - z_n(\beta r) \frac{\partial Y_{n,m}(\theta,\phi)}{\partial \theta} \hat{\phi} \right) \ldots \ldots \ldots (A3)$$

and,

$$\bar{E}_{TMnm}(\bar{r}) = \frac{1}{\sqrt{n(n+1)}} \left( \frac{n(n+1) z_n(\beta r)}{\beta r} Y_{n,m}(\theta,\phi) \hat{r} + \frac{\partial(\beta r z_n(\beta r))}{\partial(\beta r) \beta r} \frac{\partial Y_{n,m}(\theta,\phi)}{\partial \theta} \hat{\theta} + \frac{\partial(\beta r z_n(\beta r))}{\partial(\beta r) \beta r} \frac{Y_{n,m}(\theta,\phi)}{\sin(\theta)} \hat{\phi} \right) \ldots \ldots \ldots \ldots \ldots \ldots \ldots \ldots \ldots \ldots \ldots (A4)$$

Here $\{r, \theta, \phi\}$ are simply the spherical coordinates corresponding to a point $\bar{r}$. $Y_{n,m}(\theta,\phi)$ is the normalized spherical harmonics that govern the angular distribution (angular momentum) of the scattered wave whereas $z_n(r)$ represents spherical Bessel function of order $n$—denoting the radial oscillatory behavior—and appropriate type of this Bessel function need to be chosen depending on the nature of the field. For example, to represent the outward radiating wave from an arbitrary scatterer, the spherical Hankel function of type 1 is used as $z_n(r)$. On the other hand, for representing incident wave on any sphere, the spherical Bessel function of type J is used.

The quantity $n_{max}$ in equation (2) represents the maximum order to which the multipole expansion is truncated, and therefore dictates the numerical accuracy of the value of the field derived from it. The suitable value to be used for our case and the corresponding accuracy will be discussed later.

**Spherical Vector Translation:**

The response of any arbitrary set of spherical DBBs can be thought of as a multiple scattering process. The fundamental origin of this lies in the fact that the scattered wave from any arbitrary DBB *i* is scattered by any other arbitrary DBB *j*, and all these multiple scatterings events need to be self consistently accounted for in the analysis. This is achieved via the translation of spherical vector harmonics, which allows to express the scattered wave from any DBB *i* in the multipole basis centered around any other DBB *j*.

$$\bar{E}_{TEnm}(r - r_i) = \sum_{q,p} \left( \zeta_{q,p,n,m}^{i,j} E_{TEqp}^{(J)}(r - r_j) + \tau_{q,p,n,m}^{i,j} E_{TMqp}^{(J)}(r - r_j) \right) \ldots \ldots \ldots \ldots (A5)$$

$$\bar{E}_{TMnm}(r - r_i) = \sum_{q,p} \left( \zeta_{q,p,n,m}^{i,j} E_{TMqp}^{(J)}(r - r_j) + \tau_{q,p,n,m}^{i,j} E_{TEqp}^{(J)}(r - r_j) \right) \ldots \ldots \ldots \ldots (A6)$$

The radiative multipole modes comprise of the spherical Hankel function of type (1) and represent radially outward emitted wave. However, the radially outward wave originating from *i* th sphere can be also broken down in the multipole basis around the sphere *j*, leading to spherical vector harmonics with the spherical Bessel function J as the radial component (represented with the superscript (J) in equation A5 and A6). These J-type Bessel function vector harmonics represent the incident wave on the *j* th sphere that is caused by the scattered wave from *i*th sphere, in the multipole basis centered on DBB *j*. The coefficients $\zeta_{q,p,n,m}^{i,j}$ and $\tau_{q,p,n,m}^{i,j}$ are derived analytically using expansion of the Langbein function. The detailed procedure can be found in [36].

**Scattering by a Single Sphere:**

Owing to the assumed spherical symmetry of the scatterers, an incident wave of the nature $TE_{nm}^{(J)}$ will always lead to a $TE_{nm}$ type scattered wave and same applies for the TM modes as well. The conversion from the incident wave to the scattered wave can be described by a factor that is only dependent on the size, refractive index of the sphere and the wavelength of light. These are expressed as $\chi_{TEn,m}$ and $\chi_{TMn,m}$.

$$\chi_{TEn,m} = \frac{-n_{index}\, \hat{j}_n(\beta R_i)\, \hat{j}'_n(\beta_d R_i) + \hat{j}'_n(\beta R_i)\, \hat{j}_n(\beta_d R_i)}{n_{index}\, \hat{h}_n^{(1)}(\beta R_i)\, \hat{j}'_n(\beta_d R_i) - \hat{h}_n'^{(1)}(\beta R_i)\, \hat{j}_n(\beta_d R_i)} \ldots \ldots \ldots \ldots \ldots (A7)$$

$$\chi_{TMn,m} = \frac{-n_{index}\, \hat{j}'_n(\beta R_i)\, \hat{j}_n(\beta_d R_i) + \hat{j}_n(\beta R_i)\, \hat{j}'_n(\beta_d R_i)}{n_{index}\, \hat{h}_n'^{(1)}(\beta R_i)\, \hat{j}_n(\beta_d R_i) - \hat{h}_n'^{(1)}(\beta R_i)\, \hat{j}'_n(\beta_d R_i)} \ldots \ldots \ldots \ldots \ldots (A8)$$

Here $n_{index}$ denotes the refractive index of each DBB. $\beta$ and $\beta_d$ represent the electromagnetic wave-vector in the surrounding medium and in the medium of the DBB respectively.

**Analytical Method to Matrix Equation:**

With the required analytical tools discussed above, a self-consistent matrix equation is formulated. To do this, the scattered wave by the whole assembly is expressed in the form of a vector $V_{Out}$ as follows:

$$V_{out} = \begin{bmatrix} [a_{n,m}^{(1)}]_{1\times D/2} \\ [b_{n,m}^{(1)}]_{1\times D/2} \\ [a_{n,m}^{(2)}]_{1\times D/2} \\ [b_{n,m}^{(2)}]_{1\times D/2} \\ \cdot \\ \cdot \\ \cdot \\ [a_{n,m}^{(N)}]_{1\times D/2} \\ [b_{n,m}^{(N)}]_{1\times D/2} \end{bmatrix}_{1\times DN} \quad \ldots \ldots \ldots (A9)$$

where $a_{n,m}^{(i)}$ and $b_{n,m}^{(i)}$ are the coefficients in the multipole mode decomposition as shown in equation (A2). D represents the number of multipoles per DBB which can be estimated as $D = 2 \times ((n_{max} + 1)^2 - 1)$.

The spherical vector translation equations (A5) and (A6) allow us to construct a matrix that transforms the scattered wave from every DBB to incident wave. This is depicted as:

$$V_{incident} = [T]_{DN\times DN} V_{out} \ldots \ldots (A10)$$

On the other hand, equation (A7) and (A8) allow us to construct a matrix, denoted by [X], that transforms the incident wave to every DBB ($V_{incident}$) to the scattered wave ($V_{out}$). Thus, we have,

$$V_{out} = [X]_{48N\times 48N} V_{incident} \ldots \ldots (A11)$$

Equation (A10) and (A11) combined yields to a self-consistent matrix equation, given by,

$$[V_{out}] = [\chi][T][V_{out}] \ldots \ldots (A12)$$

An external source, in this scheme, can be thought of as an incident wave on the DBBs does not originate from any scattered wave from the DBBs. Thus, in the presence of an external source (as depicted in Figure A1), equation (A10) modifies to,

$$V_{incident} = [T]_{DN\times DN} V_{out} + V_{Source} \ldots \ldots (A13)$$

Thus, now equation (A12) and (A13) are combined to yield the following:

$$[V_{out}] = [\chi][T][V_{out}] + [\chi][V_{Source}] \ldots \ldots (A14)$$

Or,

$$(\mathbb{I} - [\chi][T])[V_{out}] = [\chi][V_{Source}] \ldots \ldots (A15)$$

Or,

$$[V_{out}] = (\mathbb{I} - [\chi][T])^{-1}[\chi][V_{Source}] \ldots\ldots\ldots\ldots\ldots\ldots\ldots\ldots\ldots\ldots\ldots\ldots\ldots\ldots\ldots\ldots\ldots (A16)$$

Equation (A16) shows the final matrix equation of dimension DN x DN that is solved readily using matrix inversion to obtain the scattered field for any arbitrary source.

**Test of Convergence and Error in the Numerical Analysis:**

Now we employ the Mie scattering based numerical tool discussed above to obtain the requirement on convergence of the calculated scattered field quantities. For this, a simplified case is shown in Figure A2(a), where a single spherical DBB of radius 129nm and refractive index 3.5 (representing GaAs) is excited by a plane wave source. The scattering cross section spectrum is shown in Figure A2(b), also indicating the field distribution of the few magnetic and electric multipoles of lowest orders. We note, since the field distribution for individual multipoles is calculated analytic form, the source of error in the analysis does not lie in finiteness in grid size in space and time but lies in the finiteness in the number of elements in the multipole basis that is captured by the parameter $n_{max}$ in equation (A2).

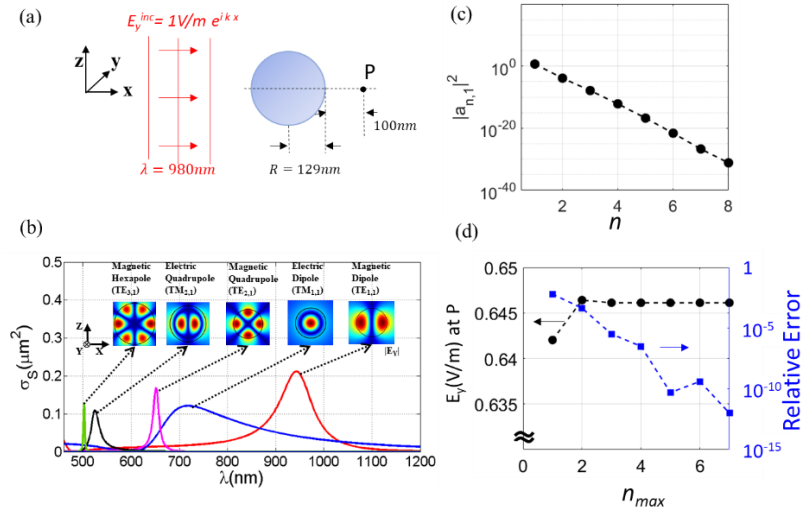

Figure A2. (a) A simple case of y polarized plane wave incident on a single GaAs DBB of size 129nm. (b) The calculated scattering cross section along with the field distribution of the first few multipole modes. (c) The multipole mode coefficients ($|a_{n,1}|$) as a function of the order n, showing an exponential decrease in the multipole mode apmplitude with increasing order. (d) The calculated $E_y$ at point P of panel (a) for increasing values of $n_{max}$, showing the convergent behavior of the multipole expansion. The relative error of the calculated Ey with respect to the asymptotic value is plotted in the blue squares.

To determine suited value of $n_{max}$, in figure A2(c), we show the magnetic multipole mode coefficients as a function of the mode order n, where an exponential drop is observed. This suggests very fast convergence of the calculated field quantities as the higher order multipole mode's contribution drops to ~$10^{-7}$, even at $n_{max} = 4$. This is further illustrated in Figure A2(d), where the calculated scattered E-field at a point P (arbitrarily chosen to be ~100nm away from the surface of the DBB) is plotted as a function $n_{max.}$—shown as the black circles. Indeed, the calculated field value rapidly converged at $n_{max}> 2$. This is verified by calculating the relative error of the calculated $E_y$ field from the asymptotic value for large n$_{max}$, shown in the blue squares in panel (d). Based on these observations, a value of $n_{max}$=4 is chosen in our analysis that ensures ~$10^{-7}$ error in the calculated E and H field quantities.